In this note we wish to demonstrate that the concept of efficiency loophole, as far as a local realistic model is concerned, is a self-defeating one since it implies that the static fields (for example, electric or gravitational) of an elementary particle must be in the form of quantized superluminal pulses. The idea, however, may still be advocated by the proponents of nonlocal realistic models since it is shown that such superluminal pulses do not contradict the facts of Lorentz transformations. As an alternative to Bell-type experiments we propose an experiment by means of which one can directly verify the validity or otherwise of the concept of efficiency loophole.
\\

It has been proposed [1] that in the Bell-type experiments the detection efficiency is an effect not only of the random errors in the analyzer and detector equipment, but is also due to a hidden physical property of the elementary particles. Thus, the result of a given measurement, say, on the spin, will yield not two, but one of the three results; $\pm\frac{\hbar}{2}$ and no detection.

This concept of no detection or no result has been used in support of a local and deterministic interpretation of quantum mechanics [2] and also as a means of achieving a sort of unification between the Copenhagen and realistic interpretations of quantum mechanics [3].

In the following we will demonstrate that, carefully considered, the idea of efficiency loophole is a problematic one and in any event it may not be advocated in support of a local realistic model.

Let us first note that the conjectured hidden physical property must in two respects be a universal one. Firstly, it must be common to all types of the elementary particles. Secondly, it must pertain to all type of physical fields. By the latter we mean the following.

Assume that a detector is designed to register the arrival of a particle by measuring its maximum electrostatic field. However, because of the hidden physical property, on occasions the particle will arrive at the detector, but the detector will fail to register its arrival. Now, we may design another type of detector. This time the detector will register the arrival of the particle by measuring its maximum gravitational field. Obviously, both types of detectors should yield the same type of statistics otherwise one will obtain different statistics for different types of interactions.

Consider a detector that is specifically designed to register the arrival of an electron by measuring its electrostatic field. If because of the hidden physical property the arrival of the electron is not registered it then follows that for a time duration $\Delta t$ the electrostatic interactions between the electron and the constituent particles of the detector were terminated. This situation, translated into concrete physical terms, means that interactions between elementary particles is an intermittent one. Also, that in contrast to the picture of an electric pole, as obtained from experiments at the macroscopic level, the Coulomb charge of an elementary particle is not uniformly distributed inside or on the surface of a sphere.

For the sake of discussion we may imagine the electron to be in the form of a spherical object, with its Coulomb charge distributed on only part of its surface area. It will now be seen that the electrostatic field of the electron would be non-isotropic unless it is assumed that the form of the charge distribution together with the internal rotational motion of the electron (as evidenced by its spin) somehow combine and present, on the average, an isotropic electrostatic field for the electron. In other words if one were to observe the electric field of an electron one would see pulses of electric field of time duration $\Delta t$, arriving at time intervals of $(1+\beta^{-1})\Delta t$ at the point of observation (fig. 1), with $\beta$ having a value less than one[1]. The value of the electric field intensity in each pulse is equal to $(1+\beta^{-1})E_r$ where, $E_r$ is the average and measurable value of the electric field.

In the classical conception of things, the electrostatic field of a charge is in the form of continuous lines emanating radially from it, and their speed of travel through space is equal to the speed of light. In the above picture, however, the electric fields are in the form of quantized pulses. Therefore, on the average they would imitate the classical field if only the speed of travel of each pulse were to be equal to $C = (1+\beta^{-1})c$. This value for $C$ is mandatory, since only in this way the Lorentz transformations will retain their form.

---

[1] In a subsequent paper to appear shortly we shall present evidence in favour of the argument that $\beta \cong \alpha$ where $\alpha \cong \dfrac{1}{137}$ is the fine-structure constant.

Obviously the value of $C$ is greater than the speed of light. Therefore, the concept of efficiency loophole can not be used in support of a local realistic interpretation of quantum mechanics. However, there is no bar in advocating the concept in support of a nonlocal realistic model where, the results of the measurements in the Bell-type experiments is now effected by the intermittent interaction and the nonlocality is attributed to the existence of the superluminal signals. The validity or otherwise of this version of nonlocality can be tested by means of the following experiment.

An electron beam is chopped so as to produce an electron pulse moving with velocity **v** in the X-direction (fig. 2). The electron pulse passes the point A midway between two metallic balls. When the electron pulse reaches the point B a voltage V is impressed between the two balls. Now, assume that electrostatic signals travel, as in the classical case, with the velocity **c**. We place the electron detector at a point C so that the time for the electrostatic field to travel from A to C will be greater than the time required for the electron pulse to travel from B to C. In this case the electron pulse will not be subjected to the dipole field. Therefore, it will not suffer a deflection and will arrive at the detector.

On the other hand if the electrostatic field is in the form of superluminal pulses then the electron pulse will suffer a deflection and will not arrive at the detector.

In principle this set-up can be used for measuring the velocity of the superluminal pulses by increasing the distance AB until the point is reached where the electron pulse eventually arrives the detector. The speed of the superluminal pulses is then obtained from the relation

$$C = \frac{AC}{BC}v \qquad (1)$$

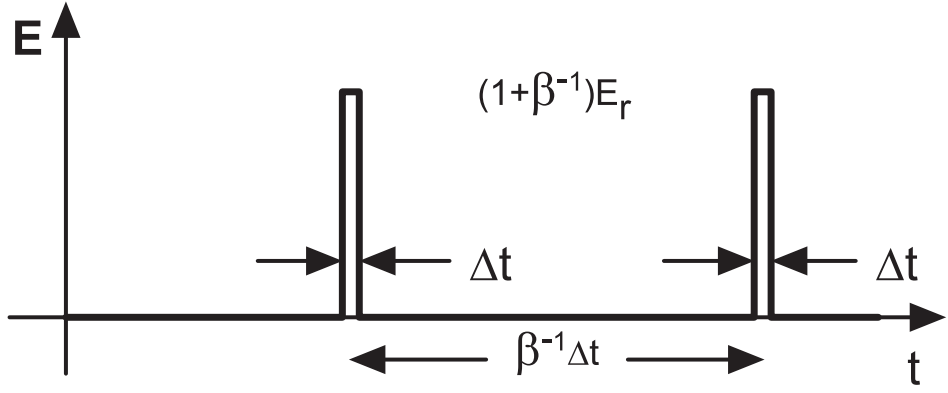

Fig. 1

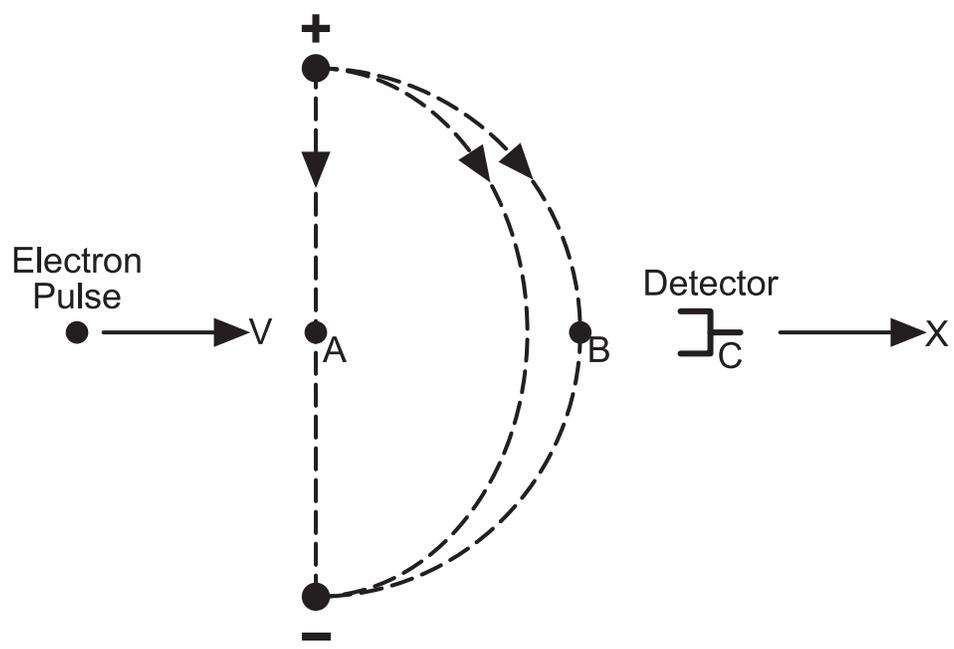

Fig. 2